\documentclass[12pt]{revtex4}

\usepackage{latexsym}
\usepackage{bbm}
\usepackage{amssymb}
\usepackage{graphicx}
\usepackage{graphics}

\newcommand{\beq}{\begin{equation}}
\newcommand{\eeq}{\end{equation}}
\newcommand{\bea}{\begin{eqnarray}}
\newcommand{\eea}{\end{eqnarray}}
\newcommand{\bra}[1]{\left\langle #1 \right\vert}
\newcommand{\ket}[1]{\left\vert #1 \right\rangle}
\newcommand{\al}{\alpha}

\newcommand{\R}{\hat{R}}
\newcommand{\F}{\cal F}
\newcommand{\kr}{\hat{E}}
\newcommand{\den}{\hat{\rho}}

\newcommand{\syn}{\hat{S}}
\newcommand{\bin}[1]{\textrm{Bin}(#1)}
\newcommand{\id}{\hat{I}_L}
\newcommand{\fid}{\cal F}

\begin{document}

\title{Combating entanglement sudden
death with non-local quantum-error correction }
\author{Isabel Sainz}
\author{Gunnar Bj\"ork}
\affiliation{School of Information and Communication Technology,
Royal Institute of Technology (KTH), Electrum 229, SE-164 40 Kista,
Sweden}

\date{\today}

\begin{abstract}
We study the possibility of preventing finite-time disentanglement
caused by dissipation by making use of {\it non-local} quantum error
correction. This is made in comparison of previous results, where
was shown that {\it local} quantum error correction can delay
disentanglement, but can also cause entanglement sudden death when
is not originally present.
\end{abstract}

\maketitle

\section{Introduction}

In the last decade substantial attention has been devoted to
entanglement because it is considered to be the physical resource on
which quantum technologies are  based \cite{nielsen}. For this
reason, entanglement dynamics have been widely studied, with the
goal of finding ways to manipulate entangled states in a practical
way, and in addition, to get a fundamental understanding of
entanglement and all its embodiments.

When interacting with a reservoir the fragile quantum states are
easily destroyed. E.g., several years ago it was shown that a pair
of initially entangled qubits can loss their entanglement in a
finite time and not asymptotically as one would naively expect
\cite{karol}. This phenomena was eventually called entanglement
sudden death \cite{esd} (ESD) and several investigations have
followed, see \cite{esd}-\cite{eberly}. Recently ESD was
demonstrated in a linear optics experiment \cite{almeida}.

Several schemes have been proposed in order to preserve fragile
quantum states when interacting with a reservoir, for example
decoherence-free subspaces \cite{dfs} and quantum error correction
(QEC) (see \cite{shor}-\cite{Fletcher}). A recent worry expressed in
\cite{eberly} is that if an entangled state suffers a ``sudden
death'' the entanglement is irrevocably lost, and therefore no QEC
scheme can help to restore the state.

Recently, using the $[4,1]$ code which encodes a single qubit into
four physical qubits and which was specifically constructed to deal
with an amplitude damping channel \cite{leung}, it has been
demonstrated that when a pair of initially entangled qubits {\it
locally} encoded with a $[4,1]\times[4,1]$ code, and subjected to
{\it local} recovery operations similar to the ones proposed in
\cite{Fletcher}, QEC can delay ESD in some cases, but may cause ESD
for certain states that will not succumb to ESD without QEC
\cite{QEC}. We would like to stress that  by {\it local} we mean
that each qubit is encoded, measured and recovered separately. In
the same work \cite{QEC}, it is mentioned that even using the {\it
non-local} code $[6,2]$ which encodes a pair of logical qubits into
six physical qubits \cite{Fletcher}, the obtained results does not
differ qualitatively from the shown in \cite{QEC}.

In this work we demonstrate this explicitly, which means that even
when the QEC protocol is {\it non-local}, in all the stages, there
is not a qualitatively difference with the {\it local} code studied
before \cite{QEC}. We also find that the success of {\it non-local}
coding and error recovery has a rather substantial state dependence.

\section{The Model}

As a simple example we will consider an amplitude damping channel,
that for one qubit can be represented in terms of the Kraus
operators :
\begin{displaymath}
\kr_0=\left(
\begin{array}{cc}
1&0\\
0&\sqrt{1-\gamma}
\end{array}\right),\qquad
\kr_1=\left(
\begin{array}{cc}
0&\sqrt{\gamma}\\
0&0
\end{array}\right),
\end{displaymath}
where $\gamma$ is the jump probability from the excited ($\ket{1}$)
to the ground ($\ket{0}$) state, $\kr_0$ is the one qubit no-jump
operator that leaves the ground state unchanged, but decreases the
excited state probability with the factor $1-\gamma$, while $\kr_1$
represents the jump operator that transforms the state $\ket{1}$
into the state $\ket{0}$ with probability $\gamma$.

To study the possibility of protecting a two qubit state and
particularly entanglement by {\it non-local} coding, where the pair
of qubits are encoded together,  and to compare it with previous
results obtained by {\it local} coding \cite{QEC}, we will use the
$[6,2]$ code introduced in \cite{Fletcher}. This code is specially
made for protecting two qubits against this kind of disturbance. It
encodes an arbitrary two-qubit pure state
\beq\ket{\varphi_0}=\cos\al\cos\delta\ket{11}+\sin\al\cos\delta
e^{i\epsilon_1}\ket{00}+\cos\beta\sin\delta
e^{i\epsilon_2}\ket{10}+\sin\beta\sin\delta
e^{i\epsilon_3}\ket{01}\nonumber\eeq into the corresponding logical
state
\bea\ket{\varphi_0}_L&=&\cos\al\cos\delta\ket{11}_L+\sin\al\cos\delta
e^{i\epsilon_1}\ket{00}_L\nonumber\\
&&+\cos\beta\sin\delta e^{i\epsilon_2}\ket{10}_L+\sin\beta\sin\delta
e^{i\epsilon_3}\ket{01}_L\nonumber,\eea were the codewords
$\ket{00}_L$, $\ket{01}_L$, $\ket{10}_L$ and $\ket{11}_L$, are given
by,\cite{Fletcher} \bea
\ket{00}_L&=&\left(\ket{000000}+\ket{111111}\right)/\sqrt{2},\label{basis00}\\
\ket{01}_L&=&\left(\ket{001001}+\ket{110110}\right)/\sqrt{2},\label{basis01}\\
\ket{10}_L&=&\left(\ket{000110}+\ket{111001}\right)/\sqrt{2},\label{basis10}\\
\ket{11}_L&=&\left(\ket{110000}+\ket{001111}\right)/\sqrt{2}.\label{basis11}\eea
Above, we have used the notation
$\ket{000000}=\ket{0}\otimes\ldots\otimes\ket{0}$, etc. It is worth
noticing that the codewords (\ref{basis00})-(\ref{basis11}) in
principle can be mapped in any way onto the physical states on the
right-hand side. With different mappings we will obtain different
results for a given state $\ket{\varphi_0}$, where the labeling of
the physical state $\left(\ket{000000}+\ket{111111}\right)/\sqrt{2}$
with the possible codewords $\ket{ij}_L$, $i,j=0,1$ is the main
cause of such differences.

We will suppose that every physical qubit is interacting with its
own environment, a situation that leads to ESD (see
\cite{karol}-\cite{eberly} and references therein). This implies
that the amplitude damping channels can be considered to be
independent, which means that the many-qubit amplitude-damping
channel will be described by the tensorial product of the
corresponding Kraus operators as in \cite{QEC}. For simplicity we
will also assume that all the qubits are damped with the same
probability $\gamma$.

As was pointed out in \cite{Fletcher}, the one physical-qubit
damping-errors are spanned by orthogonal subspaces. Nevertheless, in
this channel we have conditional evolution (the no-damping error
introduces some distortion in the original state). However, the
distortion generated by this channel in this code is of the order of
$\gamma^2$, so it satisfies the approximate QEC conditions
\cite{leung}. Therefore the no-damping evolution is
``approximately'' spanned by the codewords
(\ref{basis00})-(\ref{basis11}). Every one of the subspaces related
to no-damping and one-qubit damping are spanned by four
64-dimensional vectors
$\{{\ket{R_{k00}},\ket{R_{k01}},\ket{R_{k10}},\ket{R_{k11}}}\}$,
where $k=0$ denotes no damping, $k=1,\ldots,6$ means damping in the
first to the sixth physical qubit respectively, and the last two
subscripts ($ij$) in each $\ket{R_{kij}}$ refer to the corresponding
two-qubit state. In total we have 28 vectors spanning the no-damping
and one-qubit damping. For no damping we have the codewords
(\ref{basis00})-(\ref{basis11}), \beq
\ket{R_{000}}=\ket{00}_L,\quad\ket{R_{001}}=\ket{01}_L,\quad\ket{R_{010}}=\ket{10}_L,\quad\ket{R_{011}}=\ket{11}_L.\nonumber
\eeq Meanwhile, for the one-qubit damping one obtains \bea
\ket{R_{100}}=\ket{011111}\quad\ket{R_{101}}=\ket{010110}\quad\ket{R_{110}}=\ket{011001}\quad\ket{R_{111}}=\ket{010000}\nonumber,\\
\ket{R_{200}}=\ket{101111}\quad\ket{R_{201}}=\ket{100110}\quad\ket{R_{210}}=\ket{101001}\quad\ket{R_{211}}=\ket{100000}\nonumber,\\
\ket{R_{300}}=\ket{110111}\quad\ket{R_{301}}=\ket{000001}\quad\ket{R_{310}}=\ket{110001}\quad\ket{R_{311}}=\ket{000111}\nonumber,\\
\ket{R_{400}}=\ket{111011}\quad\ket{R_{401}}=\ket{110010}\quad\ket{R_{410}}=\ket{000010}\quad\ket{R_{411}}=\ket{001011}\nonumber,\\
\ket{R_{500}}=\ket{111101}\quad\ket{R_{501}}=\ket{110100}\quad\ket{R_{510}}=\ket{000100}\quad\ket{R_{511}}=\ket{001101}\nonumber,\\
\ket{R_{600}}=\ket{111110}\quad\ket{R_{601}}=\ket{001000}\quad\ket{R_{610}}=\ket{111000}\quad\ket{R_{611}}=\ket{001110}\nonumber.\eea
It is worth noticing that the vectors given above are 28 basis
vectors out of 64. The rest of the basis vectors can be found in the
following manner: first we should add vectors similar to
(\ref{basis00})-(\ref{basis11}), but with a minus instead of a plus.
For the rest of the basis' elements there are several ways. For
example, we can add the rest of the vectors of the computational
basis with three excitations and, we can arrange the vectors of the
computational basis with two and four excitations not considered in
(\ref{basis00})-(\ref{basis11}), in a similar way as in
(\ref{basis00})-(\ref{basis11}) and consider not only the addition,
but the substraction too. However, to list one or several of the
complete bases lies outside the scope of this paper.

In order to make the syndrome measurement more transparent, let us
introduce the following unitary transformation, that changes the
basis given above to the computational one
\beq\syn=\sum_{k=0}^{15}\sum_{i,j=0}^1\ket{ij\textrm{Bin}(k)}\bra{R_{kij}},\eeq
where the $\textrm{Bin}(k)$ is the binary representation of the
$k=0,\ldots,15$ for the last four qubits. The error syndrome
detection is completed when we measure the last four (physical)
qubits, i.e., $\bin{k}$. This allows us to detect possible errors in
a way that preserves any superposition in the two first qubits. When
measuring $\bin{k}$, if the outcome is, for example $0000$ (see
Table \ref{SynTable}), we know that no qubit was damped. If, on the
other hand, the results are $0001,0010,0011,0100,0101$ or $0110$ we
conclude that there is a jump in the qubit one to six, respectively.
It is important to notice that this assumption is only true until a
certain ``degree'' $\sim\gamma^2$, because more than one-qubit
damping could lead the same state as one qubit damping. Similarly,
some two or four-qubit damping could lead to the state
$\ket{000000}$, which is part of the no-damping basis state.

\begin{center}
\begin{table}
\caption[10pt]{Syndrome outcomes and corresponding recovery
operations. \label{SynTable}} {\begin{tabular}{|c|c|}
\hline Syndrome & Recovery operation\\
\hline $\textrm{Bin}(0)=0000$&${\R}_0=\sum_{i,j=0}^1\ket{ij}_L\bra{ij\textrm{Bin}(0)}$\\
$\textrm{Bin}(1)=0001$&${\R}_1=\sum_{i,j=0}^1\ket{ij}_L\bra{ij\textrm{Bin}(1)}$\\
$\textrm{Bin}(2)=0010$&${\R}_2=\sum_{i,j=0}^1\ket{ij}_L\bra{ij\textrm{Bin}(2)}$\\
$\textrm{Bin}(3)=0011$&${\R}_3=\sum_{i,j=0}^1\ket{ij}_L\bra{ij\textrm{Bin}(3)}$\\
$\textrm{Bin}(4)=0100$&${\R}_4=\sum_{i,j=0}^1\ket{ij}_L\bra{ij\textrm{Bin}(4)}$\\
$\textrm{Bin}(5)=0101$&${\R}_5=\sum_{i,j=0}^1\ket{ij}_L\bra{ij\textrm{Bin}(5)}$\\
$\textrm{Bin}(6)=0110$&${\R}_6=\sum_{i,j=0}^1\ket{ij}_L\bra{ij\textrm{Bin}(6)}$\\
$\textrm{Bin}(7)=0111$& Measure the first two qubits\\
$\textrm{Bin}(8)=1000$& and project onto $\id/4$\\
$\textrm{Bin}(9)=1001$&\\
$\textrm{Bin}(10)=1010$&\\
$\textrm{Bin}(11)=1011$&\\
$\textrm{Bin}(12)=1100$&\\
$\textrm{Bin}(13)=1101$&\\
$\textrm{Bin}(14)=1110$&\\
$\textrm{Bin}(15)=1111$&\\\hline
\end{tabular}}
\end{table}
\end{center}

Once that we have measured the syndrome we should apply a proper
recovery operation, summarized in Table \ref{SynTable}. If the
outcome is ``$\textrm{Bin}(k)$'' for $k=0,\ldots,6$ we will preserve
the superpositions between the two first qubits by applying the
operation $\hat{R}_k$. If we measure ``$\textrm{Bin}(k)$'' for
$k=7,\ldots,15$ we know that more than one qubit was damped, and we
cannot correct the state, but we will try to keep some features of
the original state by projecting the system to the state $\id/4$ for
reasons of symmetry, where
$\id=\ket{00}_L\bra{00}_L+\ket{01}_L\bra{01}_L+\ket{10}_L\bra{10}_L+\ket{11}_L\bra{11}_L$
is the identity in the codeword space. In this scheme, it is always
possible to reconstruct a two-qubit density matrix from the
six-qubit recovered state, because the state after recovering is in
the four-dimensional codeword sub-space, which enables us to compute
(measure) not only the exact fidelity but the pairwise entanglement
by means of the standard concurrence.

\section{Entanglement Sudden Death and Quantum Error Correction}

As was pointed out in the Introduction, ESD is present in different
scenarios for some bipartite states. In this context the so-called
$X$-states \cite{esd1} have been widely studied \cite{xstates}. They
have the property that when interacting with independent
environments, the corresponding composite density matrix preserves
the $X$ form in the computational basis. They also merit studying
because they include the Bell-states and the Werner-states.
Furthermore, among them one can find states that are subject to ESD
and states that are not, when evolving under dissipation.
Particularly, we will work with the Bell-like states of the form
\bea\label{phi}
\ket{\phi_0}=\cos\al\ket{11}+e^{i\beta}\sin\al\ket{00},\\
\label{psi}
\ket{\psi_0}=\cos\al\ket{10}+e^{i\beta}\sin\al\ket{01}.\eea It is
worth noticing that state $\ket{\phi_0}$ will succumb to ESD for
$\alpha$ such that $\vert\tan\alpha\vert<1$, this was experimentally
shown \cite{almeida}, meanwhile $\ket{\psi_0}$ will not become
disentangled for any finite amount of dissipation. In order to show
the effect of different mappings of the codeword space we will also
consider the
separable states \bea\label{zeta}\ket{\zeta_0}&=&\cos\al\ket{01}+e^{i\beta}\sin\al\ket{00},\\
\label{xi}
\ket{\xi_0}&=&\cos\al\ket{11}+e^{i\beta}\sin\al\ket{10}.\eea

We will try to protect the information carried by the states
(\ref{phi})-(\ref{xi}) by encoding these states using the codewords
given by the set of equations (\ref{basis00})-(\ref{basis11}), so
that the corresponding logical states are
\bea\label{phil}\ket{\phi_0}_L&=&\cos\al\ket{11}_L+e^{i\beta}\sin\al\ket{00}_L,\\
\label{psil}\ket{\psi_0}_L&=&\cos\al\ket{10}_L+e^{i\beta}\sin\al\ket{01}_L,\\
\label{zetal}\ket{\zeta_0}_L&=&\cos\al\ket{01}_L+e^{i\beta}\sin\al\ket{00}_L,\\
\label{xil}\ket{\xi_0}_L&=&\cos\al\ket{11}_L+e^{i\beta}\sin\al\ket{10}_L.\eea

The standard way to quantify the efficiency of quantum information
processes and protocols, particularly QEC is the fidelity, that for
an initially pure state is just the overlap between the
original/desirable state $\ket{\varphi_0}$ and the obtained one
$\den$ after dissipation and application of the considered protocol.
That is ${\fid}=\bra{\varphi_0}\den\ket{\varphi_0}$. In this case
the fidelities between the initial encoded states
(\ref{phil})-(\ref{zetal}) and the corresponding ones after
amplitude damping  modeled by the Kraus operators and QEC are given
by
\bea {\F}_{\phi}={\F}_{\zeta}&=&1-\frac{\gamma^2}{4}\left(21-9\cos2\al-\sin^22\al\cos^2\beta\right)+O(\gamma^3)\label{fidphi}\\
{\F}_{\psi}={\F}_{\xi}&=&1-\frac{\gamma^2}{4}\left(12-\sin^22\al\cos^2\beta\right)+O(\gamma^3).\label{fidpsi}
\eea  Here, the {\it non-locality} and the codeword labeling is
manifested in the fact that the fidelities when considering the
initial states $\ket{\phi_0}_L$ and $\ket{\zeta_0}_L$ are exactly
the same (not only the first terms in the series expansion given
above). This also happens for the initial states $\ket{\psi_0}_L$
and $\ket{\xi_0}_L$. This is because when considering independent
damping $\gamma$, there is no difference in the behavior between the
physical states in (\ref{basis01})-(\ref{basis11}), while if we use
the $[4,1]\times[4,1]$ code (encoded with the same labels in each
qubit), the unique logical states that will behave in the same way,
which is a desirable feature, are $\ket{01}_L$ and $\ket{10}_L$.

Disregarding this, let us study the effect of {\it non-local} coding
for a pair of qubits and the possibility of obtaining better
results, particularly protection against ESD, than with the {\it
local} coding considered in \cite{QEC}. In order to see the effect
of the proposed QEC protocol, we should compare the fidelities given
above with the fidelities given by the initial (not encoded)
corresponding states, and the fidelities obtained with the
$[4,1]\times[4,1]$ code for a pair of qubits which are given
by\cite{QEC} \bea
{\F}_{\phi_1}&=&1-\frac{\gamma^2}{2}\left(8-3\cos2\al-2\cos^22\al\right)+O(\gamma^3),\label{fidphi1}\\
{\F}_{\zeta_1}&=&1-\frac{\gamma^2}{4}\left(15-3\cos2\al-2\sin^22\al\cos^2\beta\right)+O(\gamma^3),\label{fidzeta1}\\
{\F}_{\psi_1}&=&1-\gamma^2\left(4-\cos^22\al\right)+O(\gamma^3),\label{fidpsi1}\\
{\F}_{\xi_1}&=&1-\frac{\gamma^2}{4}\left(9-3\cos2\al-2\sin^22\al\cos^2\beta\right)+O(\gamma^3),\label{fidxi1}\eea
where we use the subscript ``1'' to indicate the use of the
$[4,1]\times[4,1]$ in both the fidelities and in the (coded) states
($\ket{\varphi}_{L_1}$). For the uncoded states
(\ref{phi})-(\ref{xi}) we will similarly label the fidelities with
the subscript ``0'', and these fidelities are given by \bea
{\F}_{\phi_0}&=&1-2\gamma\cos^2\al+\gamma^2\cos^2\al,\label{fidphi0}\\
{\F}_{\zeta_0}&=&1-\gamma\cos^4\al-\frac{\gamma^2}{16}\sin^22\al+O(\gamma^3),\label{fidzeta0}\\
{\F}_{\psi_0}&=&1-\gamma,\label{fidpsi0}\\
{\F}_{\xi_0}&=&1-\frac{\gamma}{2}\left(5+2\cos2\al+\cos^22\al\right)
+\frac{\gamma^2}{8}\cos^2\al\left(3+5\cos2\al\right)+O(\gamma^3),\label{fidxi0}\eea

From Equations (\ref{fidphi})-(\ref{fidxi1}) is easy to see that
with the $[6,2]$ code, the loss of the fidelity is of the same order
as for the $[4,1]\times[4,1]$ code, in both cases decreases as
$\sim\gamma^2$, while the decrease is linear for the uncoded states
(\ref{phi})-(\ref{xi}). A first-sight difference between the $[6,2]$
code and the $[4,1]\times[4,1]$ is that the fidelity of the
entangled states in the first case depends on the parameter $\beta$
already in the second term in the series expansion for the Bell-like
states, meanwhile for the second case this dependence is weak
\cite{QEC}.

\begin{center}
\begin{figure}
\includegraphics[width=0.45\textwidth]{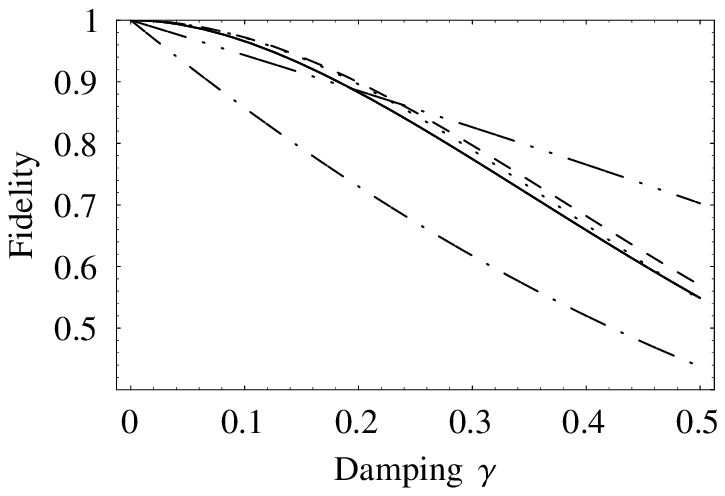}(a)
\includegraphics[width=0.45\textwidth]{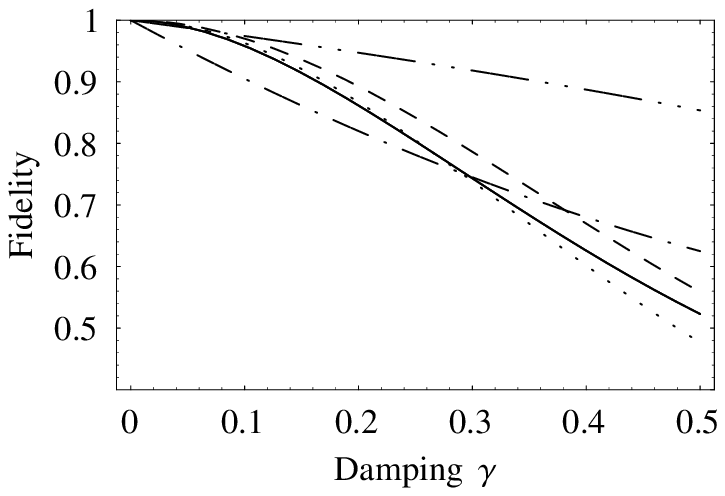}(b)
\caption[10pt]{Plot of the fidelities against the damping parameter
$\gamma$, for the state $\vert\phi\rangle$, without QEC
(dot-dashed), after recovering for the $[4,1]\times[4,1]$ code
(dotted) and the code $[6,2]$ (continuous), and for the separable
state $\vert\zeta\rangle$, without QEC (dot-dot-dashed), after
recovering for the $[4,1]\times[4,1]$ code (dashed) and the code
$[6,2]$ (continuous), when (a) $\al=\pi/6$, $\beta=0$ and (b)
$\al=\pi/4$, $\beta=0$.}\label{Figfidphi}
\end{figure}
\end{center}

\begin{center}
\begin{figure}
\includegraphics[width=0.45\textwidth]{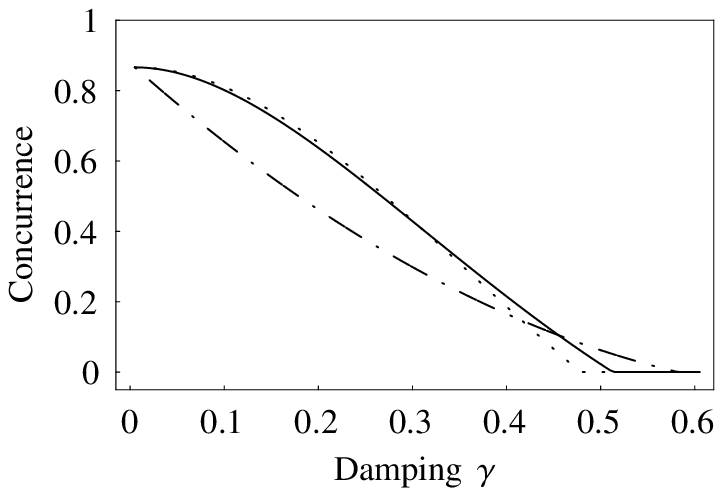}(a)
\includegraphics[width=0.45\textwidth]{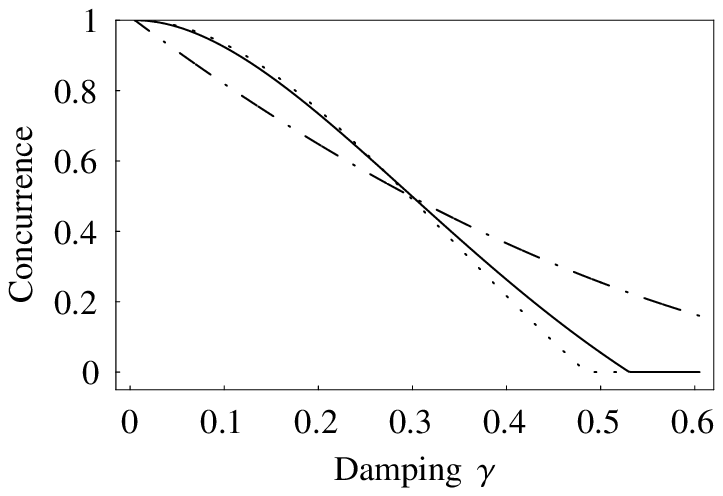}(b)
\caption[10pt]{Plot of the concurrences against the damping
parameter $\gamma$, for the state $\vert\phi\rangle$, without QEC
(dot-dashed), after recovering for the $[4,1]\times[4,1]$ code
(dotted) and the code $[6,2]$ (continuous),  when (a) $\al=\pi/6$,
$\beta=0$ and (b) $\al=\pi/4$, $\beta=0$.}\label{FigConphi}
\end{figure}
\end{center}

\begin{center}
\begin{figure}
\includegraphics[width=0.45\textwidth]{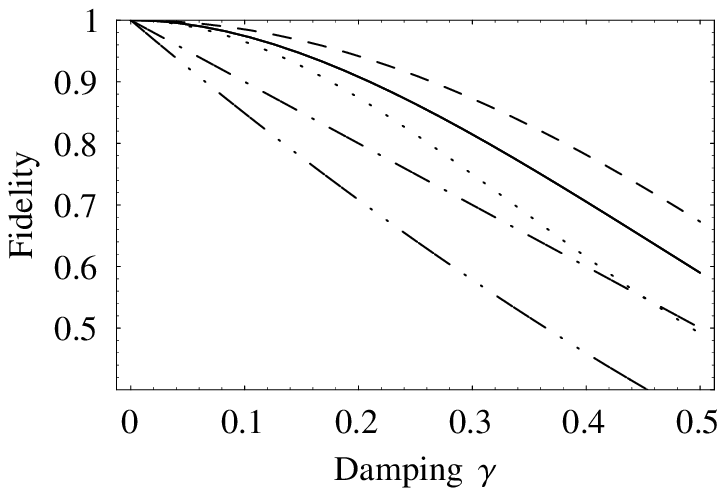}(a)
\includegraphics[width=0.45\textwidth]{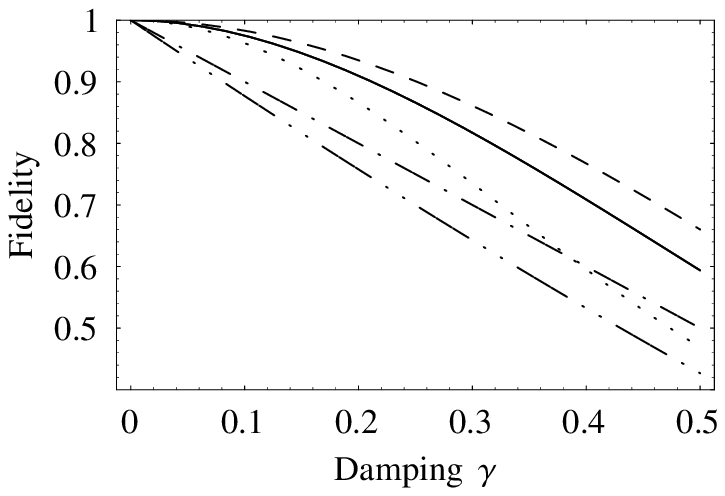}(b)
\caption[10pt]{Plot of the fidelities against the damping parameter
$\gamma$, for the state $\vert\psi\rangle$, without QEC
(dot-dashed), after recovering for the $[4,1]\times[4,1]$ code
(dotted) and the code $[6,2]$ (continuous), and for the separable
state$\vert\xi\rangle$, without QEC (dot-dot-dashed), after
recovering for the $[4,1]\times[4,1]$ code (dashed) and the code
$[6,2]$ (continuous), when (a) $\al=\pi/6$, $\beta=0$ and (b)
$\al=\pi/4$, $\beta=0$. }\label{figfidpsi}
\end{figure}
\end{center}

\begin{center}
\begin{figure}
\includegraphics[width=0.45\textwidth]{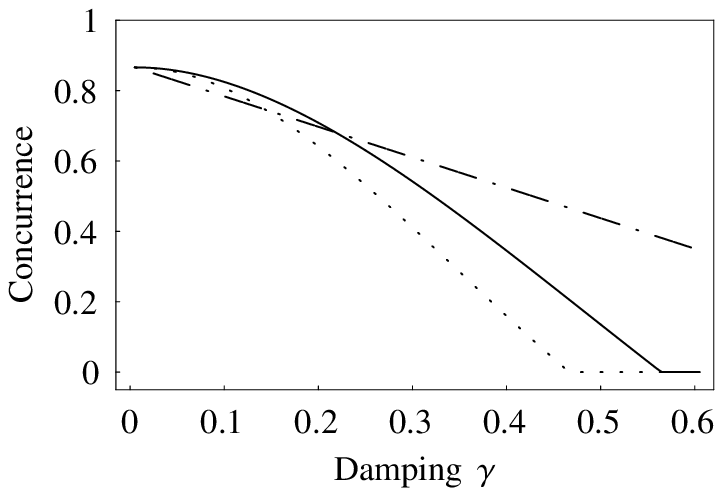}(a)
\includegraphics[width=0.45\textwidth]{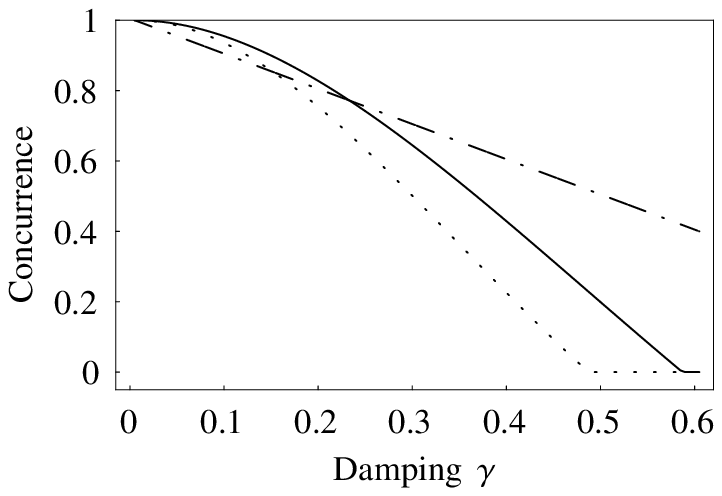}(b)
\caption[10pt]{Plot of the concurrences against the damping
parameter $\gamma$, for the state $\vert\psi\rangle$, without QEC
(dot-dashed), after recovering for the $[4,1]\times[4,1]$ code
(dotted) and the code $[6,2]$ (continuous),  when (a) $\al=\pi/6$,
$\beta=0$ and (b) $\al=\pi/4$, $\beta=0$.}\label{FigConpsi}
\end{figure}
\end{center}

Since the fidelity for the states $\ket{\phi}_L$ and $\ket{\zeta}_L$
(using the $[6,2]$ code) is the same, let us analyze its behavior
and compare it with the fidelities ${\F}_{\phi_1}$ and
${\F}_{\zeta_1}$. Even if ${\F}_{\phi}$ and ${\F}_{\zeta_1}$ depend
on the parameter $\beta$ already in the term of the order of
$\gamma^2$, the differences introduced by this parameter (which
reach their maximum when $\al=\pi/4$) are not very significant
compared with the ones introduced by $\al$, and we will disregard
this parameter, setting it to zero, in the following. In Fig.
\ref{Figfidphi} we compare the (exact) fidelities under dissipation
for the states $\ket{\phi}$ and $\ket{\zeta}$ uncoded, and for both
codes. In (a) it is shown that when $\al=\pi/6$ the fidelities do
not differ too much between the states and codes. This difference is
increased as $\al$ becomes larger, and around $\al=\pi/4$, Fig.
\ref{Figfidphi}(b), the separable state $\ket{\zeta}_{L_1}$ achieves
a higher fidelity than the Bell-state $\ket{\phi}$, and the fidelity
for this state is roughly the same for both codes. The difference
between ${\F}_{\phi}$ and ${\F}_{\zeta_1}$ remains more or less
equal until $\al=\pi/2$, but
${\F}_{\phi_1}\rightarrow{\F}_{\zeta_1}$ as $\al\rightarrow\pi/2$.
This is because the state $\ket{\zeta}_{L_1}$ has a high probability
amplitude for the physical state $\ket{0000}$, which is not evolving
(see the codewords in \cite{QEC}).
Nevertheless, in order to see how the QEC protocol is working we
should compare not only the results obtained with the different
codes, but with the fidelities without making use of any QEC. The
uncoded state $\ket{\zeta_0}$, given by Eq. (\ref{zeta}), after
damping will have such a high fidelity that both codes are
practically useless (see Fig. \ref{Figfidphi}(b)). However, in this
paper we are concerned about preventing ESD, and for the Bell-like
state $\ket{\phi}$ QEC seems to work fine in terms of the
fidelities, and we can say that for moderately large damping
parameter ($\gamma\leq0.3$) the fidelity obtained by {\it local}
coding will be higher or equal (in around $\al=\pi/4$) than the one
obtained by {\it non-local} coding. As was pointed out in
\cite{QEC}, the fidelity is insufficient to characterize the
remaining entanglement. Therefore we will study the concurrence as
well. The analytic expressions of the concurrence are cumbersome and
therefore we only present the exact, numerical plots shown in Fig.
\ref{FigConphi}, where the concurrences are almost the same for
values of $\gamma$ where it is worth to apply QEC. It is important
to notice that even for {\it non-local} QEC, the protocol will
introduce ESD when is not present in the uncoded state.  For
example, it is known that for the Bell-like states (\ref{phi}) the
entanglement disappears under dissipation only when $\gamma=1$ if
$|\cos\al|\leq|\sin\al|$.\cite{almeida} However, in Fig.
\ref{FigConphi}(b) is easy to see that this is not true when we use
either of the QEC protocols. Moreover, we do not find any
significative improvement by applying the $[6,2]$ code when
comparing with the results presented in \cite{QEC} for this states
if we consider both, the fidelity and the concurrence.
As a final step we will do a similar analysis for the states
$\ket{\psi}$ and $\ket{\xi}$, and in this case the differences
between the fidelities are more notable, ${\F}_{\xi_1}$ being the
higher and ${\F}_{\psi_1}$ being the lower in general, meanwhile for
the $[6,2]$ code the fidelity ${\F}_{\psi}={\F}_{\xi}$ is between
them. As $\al\rightarrow\pi/2$ the fidelities are roughly the same
with the value, up to second order of $\gamma$, given by
$1-3\gamma^2$, being in this case the one for {\it non-local} coding
a little higher than the other two. In Fig. \ref{figfidpsi} we plot
the exact fidelities for this states for $\beta=0$ and (a)
$\al=\pi/4$, (b) $\al=\pi/6$. We can conclude that, for this
particular Bell-like state, encoding both qubits together leads to a
significative higher fidelity than encoding separately for certain
values of $\al$ and $\beta$, this in contrast with the previous
example. In summary, we can say that in terms of the fidelity the
{\it non-local} QEC protocol presented here leads to better results
in this case.
Now let us complement our study by means of the concurrence plotted
in Fig. \ref{FigConpsi}. By this figure we see that the {\it
non-local} encoding will also leads to better results for the
concurrence. Nevertheless, as it was pointed out in our previous
work \cite{QEC} the differences between the codes is more
quantitative than qualitative, since for large values of $\gamma$
the coded state $\ket{\psi}$ loses its entanglement in a finite
time, while the uncoded state loses it asymptotically, which means
that this QEC protocol will also introduce ESD to states that
originally (uncoded) do not succumb to it. An explanation for this
behavior is given in \cite{terra}. The asymptotic state under
dissipation for the two uncoded states is the vacuum state
$\ket{00}\bra{00}$ which is a pure separable state. Therefore it
lies on the border between the sets of inseparable and separable
states. The asymptotic (logical) state obtained using QEC
corresponds to the vacuum state after applying the correction
protocol, and with the two protocols considered these will be
separable mixed states. These states lie inside the set of separable
states. Hence, a continuous evolution from an initial entangled
state ($\gamma = 0$) to the asymptotic state ($\gamma = 1$) must
hence cross the border between inseparable to separable states for a
finite amount of loss. However, for small values of $\gamma$ both
the fidelity and concurrence achieved by {\it non-local} coding are
higher than when using {\it local} coding for this Bell-like state.

\section{Conclusions}

With some examples we have studied the possibility of protecting
two-qubit Bell-like states making use of a {\it non-local} code, we
also have compared the results with previous obtained with some {\it
local} coding \cite{QEC}. We can conclude that for the Bell-like
states $\ket{\phi}$ the {\it non-local} QEC protocol presented here
does not lead to significant better results in terms of either, the
fidelity and the concurrence, while for the Bell-like states labeled
by $\ket{\psi}$ the results obtained for {\it non-local} coding are
significantly better in comparison with {\it local} coding. However,
this comparison is not completely fair because the {\it local} code,
used here, is employing eight qubits in total and is able to correct
some two-qubit errors (when each error is in different logical
qubit) whereas the {\it non-local} code, due to its shorter length,
can only correct a single error.

However, as was pointed out in \cite{QEC}, the implementation of a
$[6,2]$ code requires creation of entanglement over the same
physical distances as the entangled qubits exist over. Yet,
qualitatively the results are the same, that is, QEC can delay ESD
but it can also cause it for states that, uncoded, are not
disentangled in a finite time. This behavior can be understood in
terms of the asymptotic state formalism \cite{terra}, which suggest
that, in general, QEC will  produce ESD in the logical qubits for
any entangled state.

\section*{Acknowledgments}

The authors thank Dr. Marcelo Terra Cunha for insightful comments.
This work was supported by the Swedish Foundation for International
Cooperation in Research and Higher Education (STINT), the Swedish
Research Council (VR), and the Swedish Foundation for Strategic
Research (SSF).

\end{document}